\documentstyle[epsfig]{aipproc}

\newcommand{\bq}{\begin{equation}}
\newcommand{\eq}{\end{equation}}

\newcommand{\gsim}{\raisebox{-0.07cm}{$\, \stackrel{>}{{\scriptstyle
\sim}}\, $}}

\newcommand\GeV{\,\mbox{GeV}}
\newcommand\TeV{\,\mbox{TeV}}

\begin{document}
\begin{flushleft}
DESY 97--121 \hfill {\tt hep-ph/9707420}\\
DAPNIA SPP 97--14 \hfill June 1997\\
MSUHEP 70606, CTEQ 707
\end{flushleft}

\title{Summary of Working Group I:\\ Hadron Structure}

\author{J. Bl\"{u}mlein$^a$, J. Huston$^b$,
C. Royon$^c$ and R. Yoshida$^d$ }
\address{$^a$ DESY-Zeuthen, Platanenallee 6, D-15735 Zeuthen, Germany \\
$^b$ Department of Physics and Astronomy, Michigan State University,
East Lansing, MI 48824--1116, USA\\
$^c$ DAPNIA-SPP, Centre d'Etudes de Saclay,
F-91191 Gif-sur-Yvette Cedex, France\\
$^d$ Argonne National Laboratory, Argonne, IL 60439, USA}
\maketitle

\begin{abstract}
A summary is given on the main aspects  which were discussed by the
working group. They include new results on the deep inelastic scattering
structure functions $F_2, xF_3, F_L$ and $F_2^{c\overline{c}}$ and their
parametrizations, the measurement of the gluon density, recent theoretical
work on the small $x$ behavior  of structure functions, theoretical  and
experimental results on $\alpha_s$, the direct photon cross section, and
a discussion of the event rates in the high $p_T$ range at
Tevatron and the high $Q^2$ range at HERA, as well as possible
interpretations.
\end{abstract}

\section*{Introduction}
\noindent
Since the last International Conference on Deep Inelastic Scattering
at Rome in April 1996  various new experimental and theoretical results
have been obtained on the behavior  of the structure functions in
deeply inelastic scattering  and    related quantities.
New measurements  were performed  for $F_2(x,Q^2)$ in the low $Q^2$
domain, for $F_L(x,Q^2)$, and $F_2^{c\overline{c}}(x,Q^2)$. Furthermore
a refined analysis was carried out for the structure functions in
deep-inelastic neutrino scattering. On the theoretical side, further
investigations of the small $x$ behavior
of structure functions were carried out and various studies on the
description of their heavy flavor contributions were performed.
Several very advanced higher order calculations were carried out,
among them the 4--loop correction for the QCD $\beta$--function
in the $\overline{\rm MS}$ scheme. Different new
measurements  of the strong coupling  constant $\alpha_s(M_Z)$ were
performed.

The HERA experiments H1 and ZEUS have now accumulated a substantial
statistics in the high $Q^2$ range, $Q^2 > 10000 \GeV^2$. As a surprise
both experiments found an indication for an excess of events in this
kinematic   range over the rate predicted by the Standard Model, which
was  firstly reported in February this year.
Very intense theoretical
investigations followed in the few weeks shortly after and experimental
searches for signatures at Tevatron, which could be related, were
performed.
In the following we will give a summary on these aspects
which were discussed by the working group on
{\sf Hadron Structure}.
\section*{Nucleon Structure Function Measurements}
\noindent
Results of deep inelastic structure function measurements from NMC,
CCFR, and the HERA experiments were presented for $F_2^{ep}(x,Q^2)$,
$F_2^{\nu N}(x,Q^2)$, $xF_3^{\nu N}(x,Q^2)$,
$F_L(x,Q^2)$, and $F_2^{c\overline{c}}(x,Q^2)
$~\cite{S1:surrow,S1:meyer,S1:kabuss,S1:bernstein,S1:zomer,S1:paph1fl}.

\vspace*{1mm}
\noindent
\begin{center}
{${\sf F}_{\small \sf 2}^{\small \sf ep}{\sf(x,Q^{\small
\sf 2})}$ {\large \sf at low} ${\sf Q}^{\small\sf 2}$}
\end{center}

\vspace*{1mm}
\noindent
The ZEUS and
H1 collaborations have now measured $F_2$ down to $Q^2$ as low as
0.1 ${\rm GeV}^2$ and $x$ as low as a few times
$10^{-6}$~\cite{S1:surrow,S1:meyer}.
Figure~1  shows the HERA results, as well as
the measurement from E665, in terms of the total virtual photon-proton
cross section $\sigma (\gamma^*p)$ versus $W^2$, the center of mass
energy squared of the photon-proton system.
Data for real photoproduction ($Q^2$=0) are also shown.
The agreement between the measurements from the two HERA experiments is
rather good. The data  show a smooth transition from a slow growth of the
cross section with $W^2$ of the real photoproduction data to a fast rise
at higher $Q^2$ corresponding to the rise of $F_2$ at low $x$.
The fits to the HERA $F_2$ data using the next-to-leading order (NLO)
QCD
evolution equations have been shown to work to an unexpectedly low
value of $Q^2$ of $\sim$1 ${\rm GeV}^2$ \cite{S1:F2H1,S1:F2ZEUS}.
Also shown in Figure 1 are model predictions by
Donnachie and Landshoff~\cite{S1:DL},
based on Regge phenomenology, as well as those
of Gl\"{u}ck, Reya and Vogt\cite{S1:GRV}, based on perturbative QCD.
These give a reasonable description at the lowest $Q^2$ and $Q^2
\gtrsim$ 1 GeV$^2$, respectively.
%
%
%

\vspace*{1mm}
\begin{center}
\noindent
{${\sf F}_{\small\sf 2}^{\nu {\small
\sf  N}}{\sf (x,Q}^{\small
\sf 2}{\sf )}$ {\large \sf and}
${\sf xF}_{\small\sf 3}^{\nu {\small\sf  N}}{\sf (x,Q}^{\small
\sf 2}{\sf )}$}
\end{center}

\vspace*{1mm}
\noindent
The CCFR collaboration presented an updated analysis \cite{S1:bernstein}
of data from  neutrino scattering on iron  with an improved estimate of
quark model parameters and systematic errors. This analysis supersedes a
previous extraction of structure functions \cite{S1:ccfr} in which a
Monte Carlo technique was used to attempt to reduce the errors on the
relative calibration between the hadron and muon energies. The re-analysis
uses the calibration directly from the test beam data taken during the
course of the experiment. The net result is a change of +2.1\% in the
relative calibration and an increase in the corresponding systematic
error to 1.4\%.

\begin{center}
\mbox{\epsfig{file=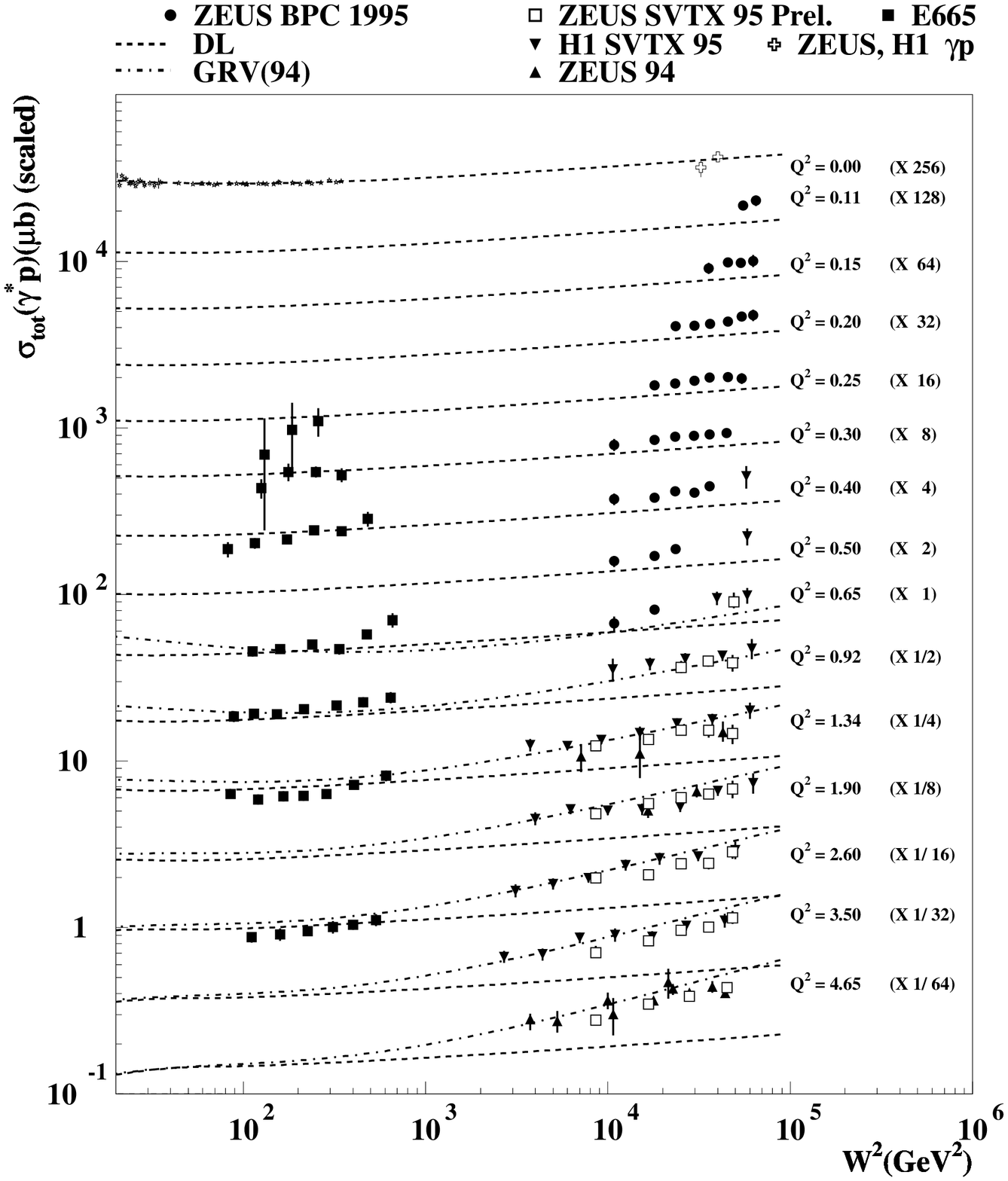,height=4in,width=4in}}
\vspace{10pt}

%
\vspace{2mm}
\noindent
\end{center}
\vspace{-7mm}
{\sf Figure~1:~Virtual photon-proton cross section measurements at
low $Q^2$, see text.}

The structure functions ${F}_{2}$ and ${xF}_{3}$ are extracted making
corrections to an isoscalar target, no strange sea, charm
mass $m_c = 0$,  and removes the QED
radiative corrections and propagator ${Q}^{2}$ dependence.
After correcting for quark-charge and heavy-target effects, the CCFR
results agree well with those from the NMC, E665, SLAC and the BCDMS
experiments for $x$
 values greater than 0.1. For $x$ values less than 0.1,
however,
the ${F}_{2}$ of CCFR
is systematically higher, with the discrepancy
reaching 15\% at a $x$
of 0.0125. These differences were also present in
the previous analysis.

The discrepancy between CCFR and NMC \cite{S1:kabuss} at low $x$
 is outside
the experimental systematic errors quoted by both groups. One possibility
is that the heavy nuclear target corrections in this region may be
different between neutrinos and charged leptons. This issue can only be
resolved with more experimental data.

\vspace*{3mm}
\noindent
\begin{center}
${\sf F}_{\small\sf L}$ {\large\sf and} {\sf R} {\large\sf  Measurements}
\end{center}

\vspace*{3mm}
\noindent
The NMC collaboration has presented \cite{S1:kabuss} their results on
the measurement of $R$~\cite{S1:paprnmc},
in the low $Q^{2}$ kinematic   domain
($1.2 \leq Q^{2} \leq 22~$GeV$^{2}$). A measurement at even lower
values of $x$ was performed by the
H1 collaboration~\cite{S1:zomer,S1:paph1fl}.
In the one photon exchange approximation, the inclusive cross section
reads
\begin{eqnarray}
\frac{d^{2} \sigma }{dxdQ^{2}} = \frac{2 \pi \alpha^{2}}{Q^{4}x}
\left( 2-2y+ \frac{y^{2}}{1+R} \right) F_{2}(x,Q^{2}),
\label{S1:E:cross}
\end{eqnarray}
with $R=F_{L}/(F_{2}-F_{L})$.
At large $y$, the weights of $F_{2}$ and $F_{L}$ in this equation become
comparable.
\begin{center}
\mbox{\epsfig{file=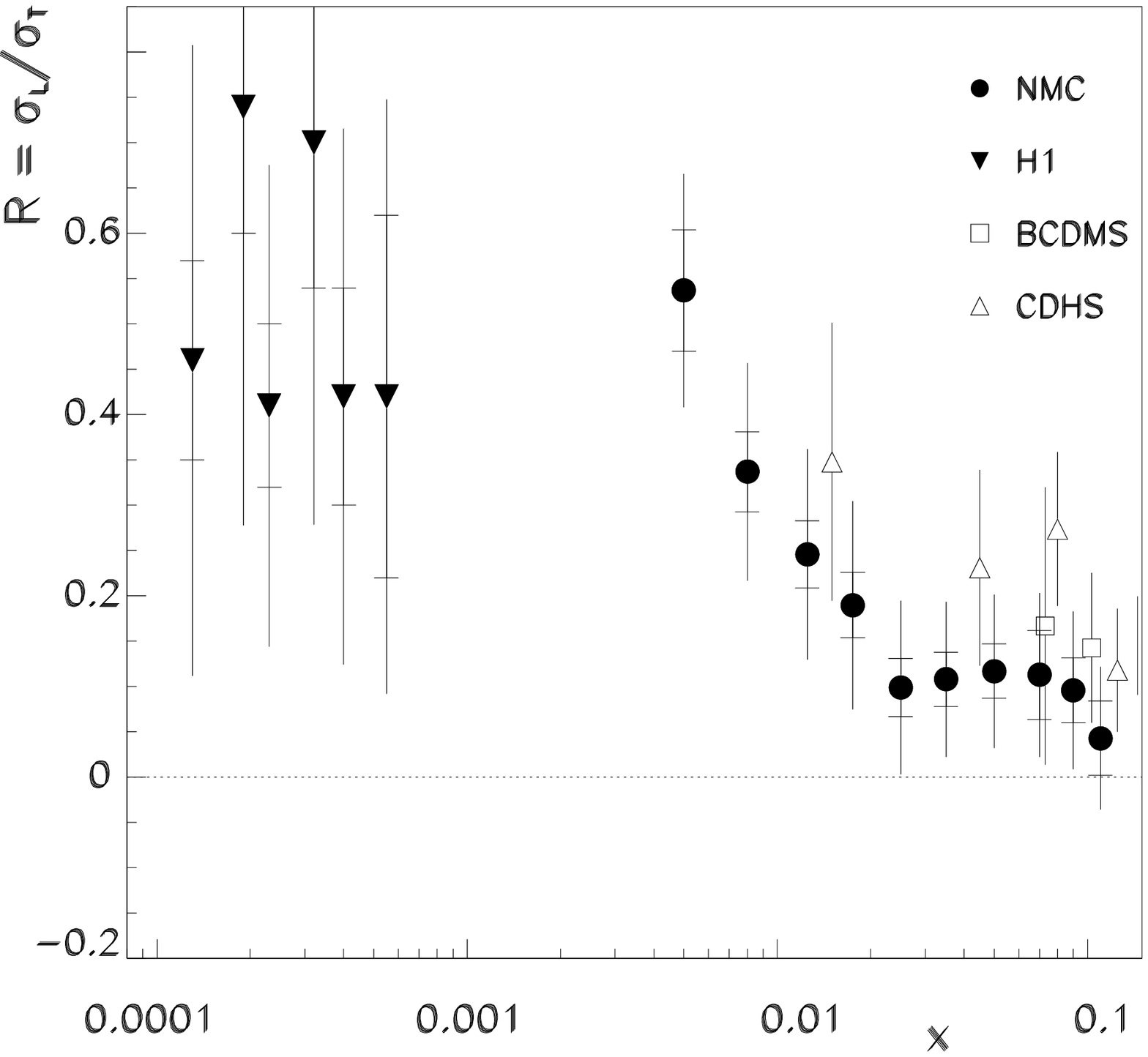,height=4in,width=4in}}
\vspace{10pt}

\vspace{-11mm}
\noindent
\end{center}
\vspace{-7mm}
\begin{center}
{\sf Figure~2:~Measurements of $R$ from NMC and H1, see text.}
\end{center}

\vspace{2mm}
\noindent
The H1 collaboration has first measured the cross section
at low $y$ where $R$  becomes negligible in eq.~(\ref{S1:E:cross}).
The structure function $F_2$ obtained in this region was then calculated
using the NLO QCD evolution equations  to larger values of $y$.
Measuring the cross section at large $y$ allows the determination of
$R$~\cite{S1:paph1fl}. The results
are at much lower $x$ than the NMC measurements  and are in good
continuity with them within the  H1 error bars,  which are still
large,~cf. Figure~2.

In order to determine
$R$ without assumptions, one  needs to measure the
cross sections at different values of $y$ fixing the values of $x$
and $Q^{2}$. This requires cross section measurements at
different values of the center of mass energy, implying a change of
beam energies. A direct change of the beam energies at
HERA has been widely discussed~\cite{S1:JBetal,S1:bauerdick},
but is in  conflict with the accumulation of integrated luminosity.
Another  method would consist in the use of  radiative events
where a real
photon is emitted in the initial state by the electron  which
reduces the beam energy  \cite{S1:krasny,S1:favart}.
However, the systematic  errors are expected to be
higher within  this method as compared to the
measurement with a lowered beam  energy.
With these  measurements, which in principle could be
performed in the near future, it may be possible to explore the
structure of $F_L$ in the low $x$ range~\cite{S1:cr}.

\vspace*{1mm}
\noindent
\begin{center}
{\large\sf Charm Contribution to} ${\sf F}_{\small\sf 2}
{(\sf x,Q}^{\small\sf 2}{\sf )}$
\end{center}

\vspace*{1mm}
\noindent
The data collected by the ZEUS and H1 experiments has allowed
a measurement of
the charm component to the proton structure function $F_{2}^{c \bar{c}}$
\cite{S1:h1fc,S1:roldan,S1:zomer}.
Compared to the previous EMC measurements, the
1994 HERA data allowed an extension  of the kinematic   domain by two orders
of
magnitude (down to $x \sim 3 \cdot 10^{-4}$ for
$Q^{2} \sim 7$ GeV$^{2}$) and
show a steep rise of $F_{2}^{c \bar{c}}$ with decreasing $x$. The new
preliminary 1995 data shown by the ZEUS collaboration reach even
lower values of $Q^{2}$ and $x$ ($Q^{2} \sim 3$ GeV$^{2}$ and
$x \sim 10^{-4}$). The H1 and ZEUS data are in good agreement as
illustrated in Figure~3. These data also agree with the results
of the NLO QCD fits performed by both collaborations. The charm treatment
\cite{S1:botje,S1:zomer} in the NLO fits is the same for both
experiments describing $F_2^{c\overline{c}}(x,Q^2)$ by the
boson--gluon
fusion process in NLO~\cite{S1:HEANLO} assuming three light flavors.
It should be noted that the dominant theoretical
uncertainty   in the $F_{2}^{c \bar{c}}$ measurement arises from the
unknown charm mass.

\begin{center}
\mbox{\epsfig{file=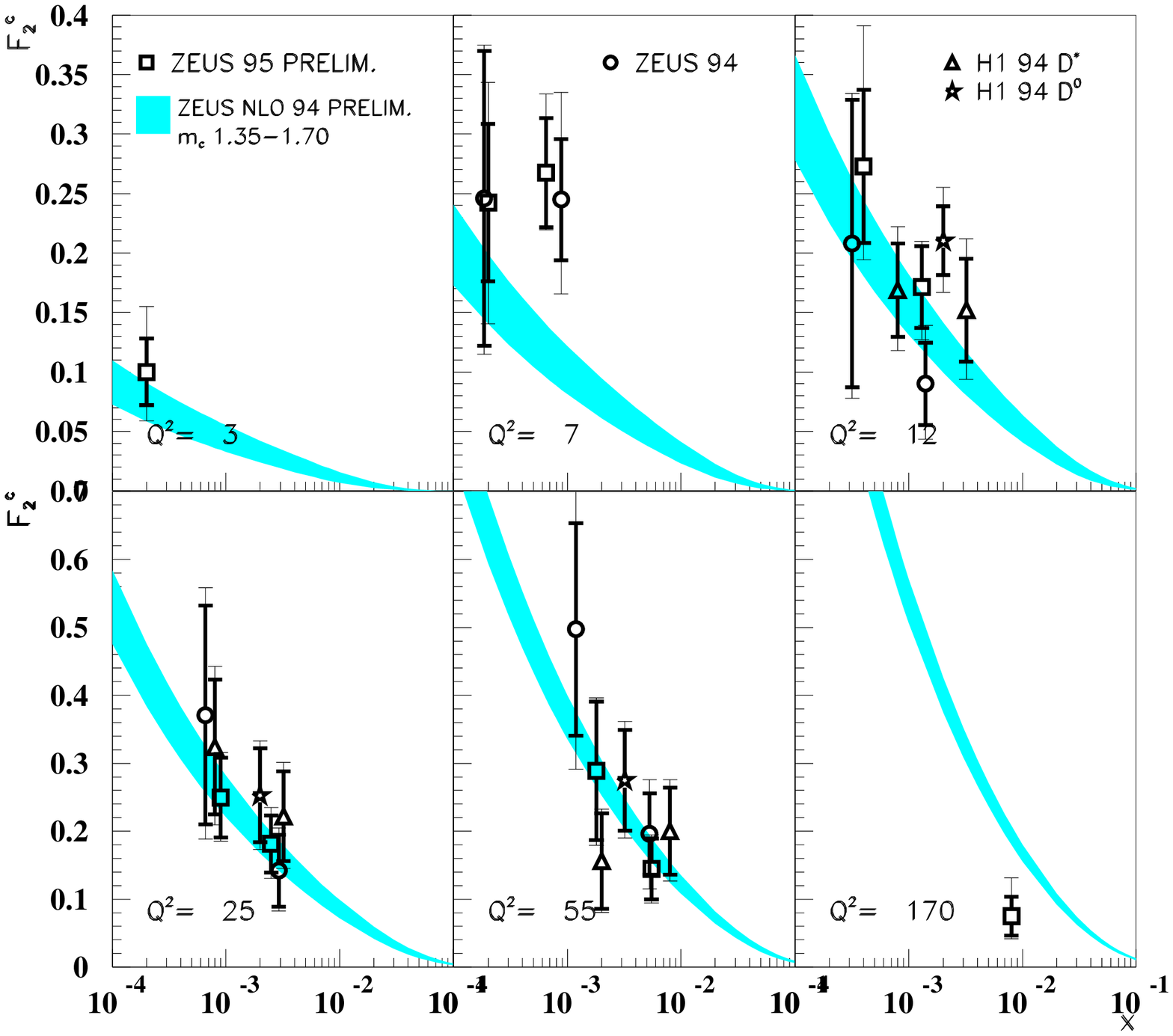,height=4.6in,width=4.4in}}
\vspace{10pt}

\vspace{2mm}
\noindent
\end{center}
\vspace{-7mm}
{\sf Figure~3:~Measurements of $F_2^{c\overline{c}}(x,Q^2)$
from ZEUS and H1, see text.}


\vspace*{1mm}
\begin{center}
\noindent
{\large\sf Gluon Extraction from Structure Functions}
\end{center}

\vspace*{1mm}
\noindent
The gluon density can be measured from the scaling violations of the
structure function $F_2(x,Q^2)$.
The results of  QCD fits  in NLO
from H1 and ZEUS are shown  in Figure~4.
The error bands due to the statistical and systematic errors are also
shown.
Both of these fits use the fixed flavor number scheme
(see above) for handling the charm quark.
In the case of the H1 analysis, NMC and BCDMS data along with their
own data are used in the fit.  For the ZEUS fit,
only NMC and ZEUS data are used.

The form of the parametrizations for the singlet, non--singlet and gluon
distributions are somewhat different between the ZEUS and H1 fits
\cite{S1:botje,S1:zomer,S1:F2H1}. The two results agree within their
errors but the H1 fit is systematically higher than the ZEUS fit.
The prospects to further constrain the gluon density in the future
are discussed below~\cite{S1:botje}.

\begin{center}
\mbox{\epsfig{file=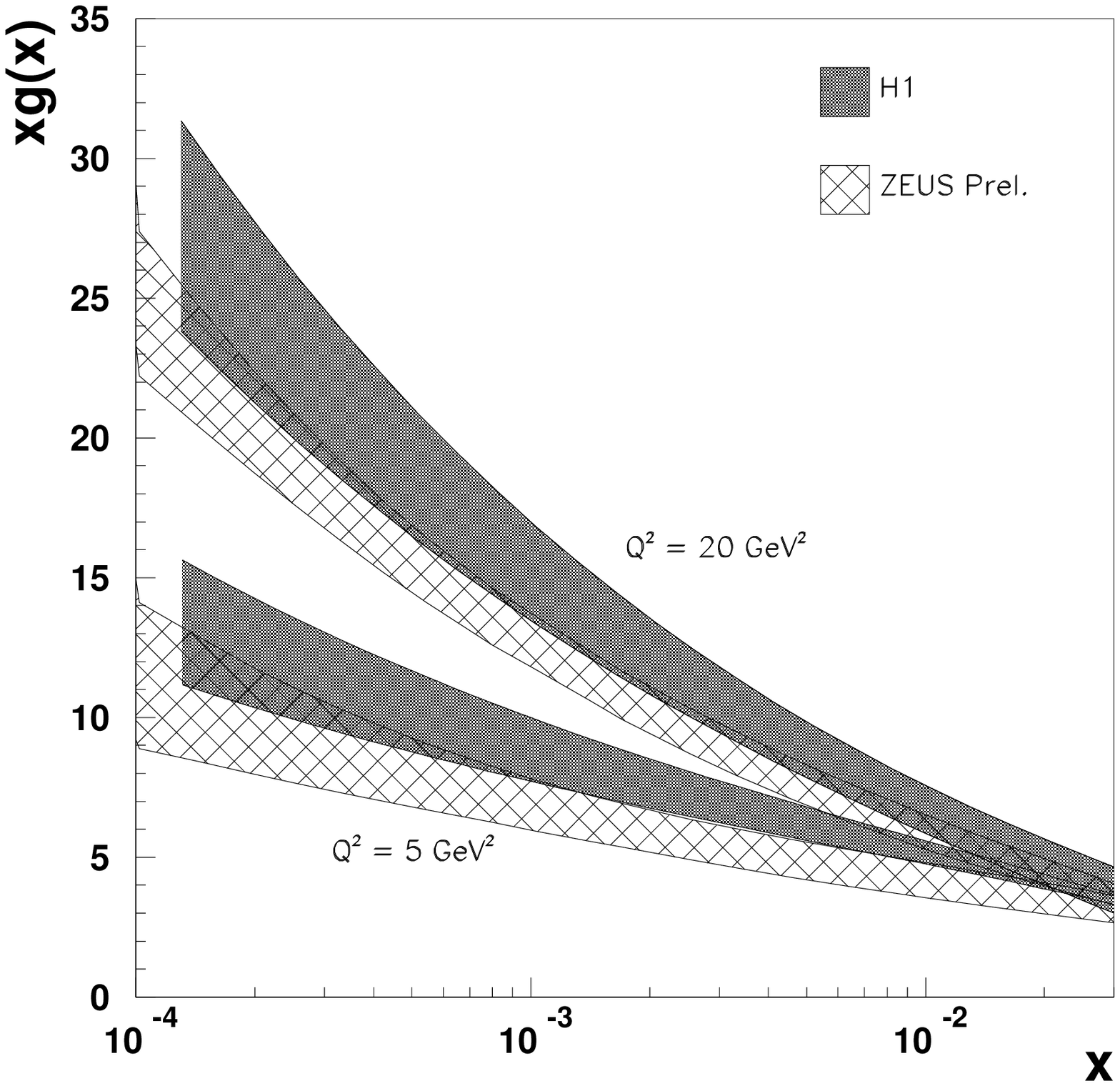,height=4.in,width=4in}}
\vspace{10pt}

\vspace{2mm}
\noindent
\end{center}
\vspace{-7mm}
{\sf Figure~4:~Gluon density distributions from H1 and ZEUS NLO QCD fits
to their 1994 $F_2$ data, see text.}
\section*{Parton Parametrizations and Heavy Flavor}
\noindent
Different updates of the parton parametrizations were reported during
this workshop~\cite{S1:cteq4,S1:MRRS} which extend earlier analyses
accounting for more dynamical effects in the heavy flavor sector.
Similar aspects were considered before as well in~\cite{S1:GRV} applying
the coefficient functions~\cite{S1:HEANLO} in a scenario based on three
light flavors. During the last year more theoretical calculations of
the heavy flavor contributions to both  the structure functions in
neutral and charged current deep inelastic scattering were
performed~\cite{S1:LT,S1:VN1,S1:VN2} and others are in
progress~\cite{S1:SO}. The data on $F^{c\bar{c}}_2(x,Q^2)$ will
further improve
as more luminosity is collected. This will allow for more detailed
comparisons between the still somewhat different theoretical predictions
in the future.

For the small $x$ range at HERA, an  effective parametrization of the
structure function $F_2^{ep}(x,Q^2)$ was derived~\cite{S1:DH}
\begin{equation}
F_2(x,Q^2) = m  \log \left (\frac{x_0}{x} \right ) \log \left (
1 + \frac{Q^2}{Q^2_0} \right ),
\end{equation}
which updates an earlier analysis~\cite{S1:BH}. This representation
depends only on three parameters, $m =  0.455$, $x_0 = 0.04$
 and $Q^2_0 = 0.55 \GeV^2$, in the range
$x < 0.005$ and the whole range of $Q^2$.
$F_2^{ep}$ exhibits a {\it logarithmic} $x$ dependence,
and so far no power behavior is indicated.

In another contribution~\cite{S1:BKW}, a model for the low $Q^2$ behavior
of $F_2^{ep}$ was presented. This model aims on a description of the
transition from the deep-inelastic to the real photon region and
describes very well part of the data. Still an improvement is needed to
match also the very recent HERA data at
$Q^2 \approx 0$~\cite{S1:surrow,S1:meyer}.
\section*{Structure Functions at Small $x$}
\noindent
A unified BFKL-GLAP approach for the deep inelastic
structure functions was presented by Kwiecinski et al.~\cite{S1:KMS}.
Among various possibilities of such a description one well-known
way consists
in relating the gluon-induced parts of the structure functions to the
un-integrated
gluon density $f_g(z,k_T^2)$. In  earlier studies, cf. e.g.
\cite{S1:AKMS}, however,
problems were encountered in the infrared region,
$k_T^2 \rightarrow 0$,
                      requesting cut--off procedures with a strong
parametric dependence. A solution to this problem was given in
ref.~\cite{S1:JB93}  where the representation
\begin{eqnarray}
F_{2,L}^g(x,Q^2) &=& \int_x^1 \frac{dz}{z} \int_0^{k_T^{2, max}} \frac{
d k_T^2}{k_T^2} \widehat{F}_{2,L}^g(z, k_T^2, Q^2) f_g\left (\frac{x}
{z}, Q^2 \right ) \nonumber\\
&=& \int_x^1 \frac{dz}{z}
\widehat{F}_{2,L}^g(z, 0, Q^2) G\left (\frac{x}
{z}, Q^2_0 \right ) \nonumber\\
&+& \int_x^1 \frac{dz}{z}
                        \int_{Q_0^2}^{k_T^{2, max}} \frac{
d k_T^2}{k_T^2} \widehat{F}_{2,L}^g(z, k_T^2, Q^2) f_g\left (\frac{x}
{z}, Q^2 \right ) + O \left( \frac{Q_0^2}{Q^2} \right)~.
\label{E:S1:REPJB}
\end{eqnarray}
was used.
For $F_2$ the collinear singularity has to be subtracted
as in the case of collinear factorization. If one assumes $Q^2 \gg
Q_0^2$ the dependence of $F_{2,L}(x,Q^2)$ in
eq.~(\ref{E:S1:REPJB})
on $Q_0^2$ is very small, cf. also~\cite{S1:JBFL}.
This description was applied in \cite{S1:KMS} representing the
un-integrated gluon density by
\begin{eqnarray}
f_g(x,k^2) &=&
 \widetilde{f}_g^{(0)}(x,k^2) + \overline{\alpha}_s(k^2)
\left[k^2 {\bf L}[f_g, k_0^2] \right]  +
\left ( \frac{x}{6} P_{gg}(x) - 1 \right ) \nonumber\\ &\otimes&
\int_{k_0^2}^{k^2} \frac{dk'^2}{k'^2} f_g(x,k'^2) +
\frac{\alpha_s(k^2)}{2 \pi} \int_x^1 dz P_{gq}(z) \Sigma\left (\frac{x}
{z}, K^2\right).
\end{eqnarray}
Here ${\bf L}$ denotes the Lipatov kernel with a lower cut-off
and $\widetilde{f}_g^{(0)}$ is the  modified
inhomogeneous term. A good fit to the $F_2$ data
was obtained using a flat input distribution for the gluon,
cf.~\cite{S1:KMS}.

In the conventional approaches, the
QCD evolution equations are considered
for parton densities
\begin{equation}
\frac{d}{d \log Q^2} \left (\begin{array}{c} \Sigma(N,Q^2)\\
G(N,Q^2) \end{array} \right ) = \left (\begin{array}{cc} P_{qq} & P_{qG}
\\ P_{Gq} & P_{GG} \end{array} \right )(N, \alpha_s) \otimes
\left (\begin{array}{c} \Sigma(N,Q^2)\\
G(N,Q^2) \end{array} \right ),
\label{S1:E:SG}
\end{equation}
and are thus renormalization and factorization scheme dependent. Already
in refs.~\cite{S1:SI} the study of evolution equations for observables
has been proposed. These equations are scheme independent and their
accuracy is only determined by the order in $\alpha_s$ which is
accounted for. Recently this approach was followed
in refs.~\cite{S1:CAT,S1:THO} choosing $F_2$ and $F_L$ as the
observables and
including the LO small $x$ resummed anomalous dimensions into the
analysis. Eqs.~(\ref{S1:E:SG}) may be transformed into evolution
equations for $F_2$ and $F_L$ eliminating the singlet and gluon
distributions by using
\begin{eqnarray}
F_2^S &=& c_2^S \Sigma + c_2^g G          \nonumber\\
F_L^S &=& c_L^S \Sigma + c_L^g G,
\end{eqnarray}
from which
\begin{equation}
\frac{d}{d \log Q^2} \left (\begin{array}{c} F_2^S(N,Q^2)\\
\hat{F}_L^S(N,Q^2) \end{array} \right ) = \left ( \begin{array}{cc}
 P_{22} &
P_{2L} \\ P_{L2} & P_{LL} \end{array} \right )
(N,\alpha_s)  \otimes
\left (\begin{array}{c} F_2^S(N,Q^2)\\
\hat{F}_L^S(N,Q^2) \end{array} \right )
\end{equation}
is obtained, where
$\hat{F}_L^S = (2 \pi/\alpha_s) F_L^S$, and  $N$ denotes
Mellin index.
As a result of the  LO + LO $\log(1/x)$--study~\cite{S1:THO}
 a lower $\chi^2$ value
was found than in different NLO analyses of the structure function
$F_2(x,Q^2)$. This applies particularly to the range $x \leq 0.1$,
whereas for $x > 0.1$ the     $\chi^2$
was somewhat larger than in the NLO analysis.

Extending earlier analyses~\cite{S1:BV1} the effect of the $N_f$ terms
in the small $x$
resummed anomalous dimension  $\gamma_{gg}(N)$, which was recently
calculated in ref.~\cite{S1:CC}, was investigated for the evolution
of $F_2(x,Q^2)$ in \cite{S1:BV}. In the ${\rm DIS}-\overline{\rm MS}$
scheme the contribution to $F_2(x,Q^2)$ is positive but suppressed
by at least a factor of 30 w.r.t. the value of $F_2(x,Q^2)$ accounting
for the resummed quark anomalous dimensions in NLO.
The small $x$ resummed contributions for $F_L(x,Q^2)$ and
$F_2^{\gamma}(x,Q^2)$
were also studied in \cite{S1:BV1}. For
$F_2^{\gamma}(x,Q^2)$
the resummation effect on the photon-specific
inhomogeneous solution originates solely  from  the resummed homogeneous
evolution operator. Since the input densities are harder if compared
to the proton case
 the small $x$
resummed terms emerge already in the range $x \sim 10^{-2}$.
\section*{$\alpha_s$ in Deep Inelastic Scattering}
\noindent
One of the most important theoretical achievements reported at this
conference is the 4--loop calculation of the $\beta$ function of QCD
in the $\overline{\rm MS}$ scheme~\cite{S1:RIT1}, which determines the
evolution equation for $a_s = \alpha_s/(4\pi)$
\begin{equation}
\label{S1:E:AL}
\frac{\partial a_s}{\partial \ln \mu^2} = - \beta_0 a_s^2
- \beta_1 a_s^3 -\beta_2 a_s^4 - \beta_3 a_s^5 + O(a_s^6)~.
\end{equation}
Herewith a further
step on the way to put QCD to the ultimate test has been performed. As
already in the case of the 2--loop result~\cite{S1:TWLO} the calculation
of the renormalized strong coupling is most easily carried out 
by considering
the renormalization of the ghost-ghost-gluon vertex,
\begin{equation}
g_{s0} = \frac{\widetilde{Z}_1}{\widetilde{Z}_3} \frac{1}{\sqrt{Z_3}}
g_s~.
\end{equation}
The calculation      still involves
30834 vertex and 21128 self-energy diagrams.
For $SU(3)_c$, where  $C_A = 3, C_F = 4/3, T_R = 1/2$,
the expansion coefficients in eq.~(\ref{S1:E:AL}) read
\begin{eqnarray}
\beta_0 &=& \frac{1}{4} \left [11 - \frac{2}{3} N_f \right]
\\
\beta_1 &=& \frac{1}{16} \left [102 - \frac{38}{3} N_f \right]
\\
\beta_2 &=& \frac{1}{64} \left [ \frac{2857}{2} - \frac{5033}{18} N_f
+ \frac{325}{54} N_f^2 \right]\\
\beta_3 &=& \frac{1}{256} \left [ \left(
\frac{149753}{6} + 3564 \zeta_3 \right)  - \left (\frac{1078361}{162}
+ \frac{6508}{27} \zeta_3 \right) N_f \right.
\nonumber\\
 & &~~~~~+ \left.
\left (
\frac{50065}{162} +
 \frac{6472}{81} \zeta_3 \right) N_f^2 + \frac{1093}{729} N_f^3 \right]~.
\end{eqnarray}
In 4--loop order new color coefficients do emerge, as well as
$\beta_3$ turns out to be
transcendental in the $\overline{\rm MS}$ scheme, unlike the case of the
lower coefficients. Recently  an attempt has been made to predict
$\beta_3$ on the basis of the lower expansion coefficients and a known
constraint applying
 the Pade approximation~\cite{S1:JE}.
The latter result differs from the complete calculation, which is partly
caused by the emergence of new color factors.

New results were also presented on the 3--loop corrections of the
Ellis-Jaffe sum rule~\cite{S1:RIT3}
\begin{eqnarray}
\int_0^1 dx g_1^{p(n)}(x) &=&
\left [ 1 - \left ( \frac{\alpha_s}{\pi} \right )
- 3.583 \left ( \frac{\alpha_s}{\pi} \right )^2
- 20.215 \left ( \frac{\alpha_s}{\pi} \right )^3 \right ]
\nonumber\\
& & \times
\left[ \pm \frac{1}{12}|g_A| + \frac{1}{36} a_8 \right ] \nonumber\\
&+& \left [1 - 0.333 \left (   \frac{\alpha_s}{\pi} \right )
- 0.550 \left ( \frac{\alpha_s}{\pi} \right )^2
         - 4.447 \left ( \frac{\alpha_s}{\pi} \right )^3 \right ]
\frac{\hat{a}_0}{9}~,
\end{eqnarray}
where $|g_A| = \Delta u - \Delta d, a_8 = \Delta u + \Delta d -
2 \Delta s$, $a_0 = \Delta u + \Delta d +
\Delta s$,
and $\hat{a}_0 = \exp[- \int^{a_s(\mu^2)} d a_s'
\gamma^s(a_s')/\beta(a_s')] a_0(\mu^2)$.
They  may become important in later QCD analyses, when much more
precise measurements of $g_1^{p(n)}(x,Q^2)$ will be
available in the future.
\begin{center}
\mbox{\epsfig{file=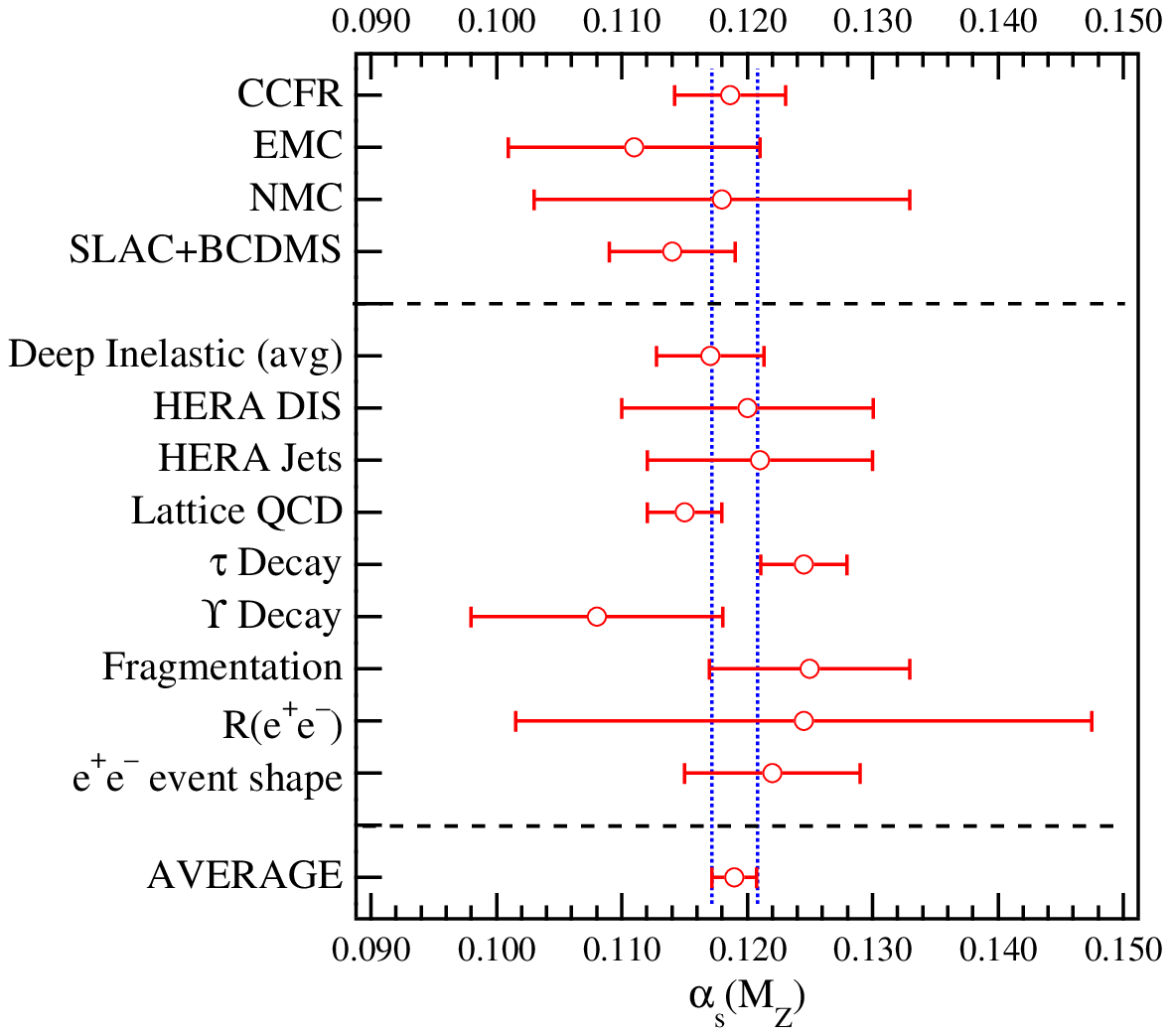,width=12cm}}
\vspace{10pt}

\vspace{2mm}
\noindent
\end{center}
\vspace{-7mm}
\begin{center}
{\sf Figure~5~: A comparison of different measurements of
$\alpha_s(M_Z^2)$, \cite{S1:ACCFR}.}
\end{center}

\vspace*{2mm}
\noindent
In a recent QCD analysis the CCFR collaboration has performed a
measurement of the strong coupling constant~\cite{S1:ACCFR} obtaining
\begin{equation}
\label{S1:E:accf}
\alpha_s^{\rm NLO}(M_Z^2) = 0.119 \pm 0.002 ({\rm exp}) \pm 0.001 ({\rm
HT}) \pm 0.004 ({\rm scale})~.
\end{equation}
This value is larger than that obtained in an
earlier measurement,
$\alpha_s^{\rm NLO}(M_Z^2) = 0.111 \pm .002 \pm .003$,
by the same experiment
and other values measured in different deep
inelastic scattering experiments, though being
compatible within the
experimental errors, see Figure~5. The largest theoretical uncertainty is due to
the renormalization and factorization scale, whereas uncertainties
due to higher twist effects (HT) are estimated to contribute only
marginally.
The NuTeV neutrino experiment, which is
currently running at the Tevatron, will
allow a number of improvements to the analysis. There is a sign-selected
beam,
which should increase anti-neutrino statistics,
and a continuous test beam which will
allow a better determination of the hadron and muon energy calibrations
and
resolutions,
currently being the largest experimental contributions to the
systematic uncertainty.

$\alpha_s(M_Z^2)$ was also determined  in a
NLO  QCD analysis of the polarized structure function
$g_1(x,Q^2)$,~\cite{S1:RID}, yielding
\begin{equation}
\alpha_s^{\rm NLO}(M_Z^2) = 0.120 \left.
\begin{array}{l} +0.004 \\ -0.005 \end{array} \right|_{\rm exp}
\left.
\begin{array}{l} +0.009 \\ -0.006 \end{array} \right|_{\rm thy}~.
\end{equation}

Further determinations of $\alpha_s$ were reported using the
GLS sum rule
\begin{equation}
\int_0^1 dx F_3(x,Q^2) = 3 \left [ 1 - \frac{\alpha_s}{\pi} - a
\left ( \frac{\alpha_s}{\pi}\right )^2 - b
\left ( \frac{\alpha_s}{\pi}\right )^3 \right]
 - \Delta{\rm HT},
\end{equation}
and the polarized Bjorken sum rule
\begin{equation}
\label{S1:E:bsr}
\int_0^1 dx \left [ g_1^p(x,Q^2) - g_1^n(x,Q^2) \right] = \frac{|g_A|}{6}
 \left [ 1 - \frac{\alpha_s}{\pi} - 3.583 \left( \frac{\alpha_s}
{\pi} \right )^2 - 20.215 \left ( \frac{\alpha_s}{\pi} \right )^3
\right].
\end{equation}
As a (preliminary) result  CCFR obtained~\cite{S1:ACCFR}
\begin{equation}
\alpha_s^{\rm NLO}(M_Z^2) = 0.112 \left.
\begin{array}{c} +0.004 \\ -0.005 \end{array}\right|_{\rm stat}
\left.
\begin{array}{c} +0.006 \\ -0.005 \end{array}\right|_{\rm syst}
\left.
\begin{array}{c} +0.004 \\ -0.005 \end{array}\right|_{\rm HT}
\pm 0.008~({\rm Model})
\end{equation}
with a lower central value than found  in the global QCD analysis,
eq.~(\ref{S1:E:accf}).

>From the Bjorken sum rule, eq.~(\ref{S1:E:bsr}), $\alpha_s(M_Z^2)$ was
determined in \cite{S1:RID} as
\begin{equation}
\alpha_s^{\rm NLO}(M_Z^2) = 0.118
 \begin{array}{c} +0.010 \\ -0.024 \end{array}
\end{equation}
with larger errors than in the QCD analysis of $g_1(x,Q^2)$ itself,
mainly due to the uncertainty in the low $x$ extrapolation.

The future
prospects to measure $\alpha_s$ from the scaling violations of
the structure function $F_2^{ep}(x,Q^2)$ were also discussed in a
contribution to this workshop~\cite{S1:BOT,S1:botje}.
In a detailed numerical
comparison four independent NLO evolution codes were found to agree
by better than $\pm 0.05\%$ for the kinematic range of
HERA~\cite{S1:COD}. Due to this the algorithmic uncertainty in the
measurement of $\alpha_s$ is well under control.

\begin{center}
\mbox{\epsfig{file=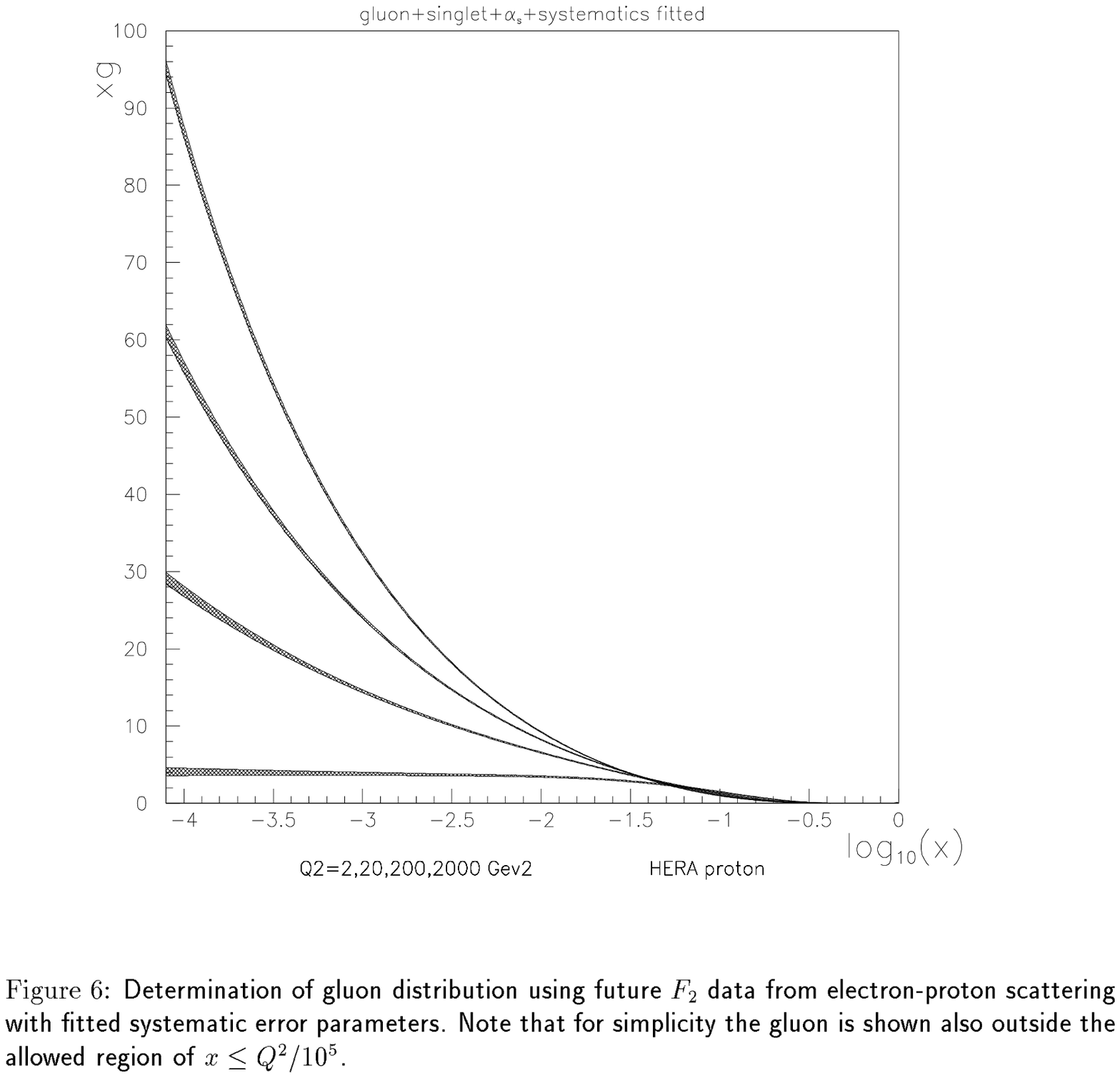,bbllx=80pt,bblly=320pt,bburx=485pt,bbury=727pt,height=9.5cm,clip=}}

\vspace{2mm}
\noindent
\end{center}
\vspace{-7mm}
{\sf Figure~6:~The gluon distribution versus $x$ for different
values of $Q^2$ obtained from a fit to simulated high energy HERA
proton data. The bands show the total error.}

\vspace{2mm}
\noindent
For an integrated luminosity of
${\cal L} = 500 pb^{-1}$
$\alpha_s(M^2_Z)$ may be measured in the
HERA range with  experimental errors of
\begin{eqnarray}
\delta \alpha_s^{\rm NLO}(M_Z^2) &=&
 0.0025 ... 0.0035~~~{\rm (systematics~fixed)} \\
\delta \alpha_s^{\rm NLO}(M_Z^2) &=&
 0.0015 ... 0.0020~~~{\rm (systematics~fitted)}~.
\end{eqnarray}
A considerable improvement can be obtained fitting systematic effects.
The constraints which can be derived for the gluon density in the latter
case are illustrated in figure~2 for an integrated luminosity of
${\cal L} = 500 pb^{-1}$.

The theoretical errors of different kind for the $\alpha_s$ measurement
were analyzed in \cite{S1:ATHEO}.
The largest contributions are due to the use of
NLO relations for $\alpha_s$  differing in NNLO,
\begin{equation}
\Delta \alpha_s^{\rm NLO}(M_Z^2) = 0.003~~~{\rm (representation~of
}~\alpha_s)
,
\end{equation}
and the renormalization and factorization scale uncertainties,
\begin{equation}
\Delta \alpha_s^{\rm NLO}(M_Z^2) = \pm 0.005~({\rm ren.)} \pm 0.003
~({\rm fact.})~~~~~(Q^2 > 50~\GeV^2).
\end{equation}
Because these errors are larger than the expected experimental errors
one would wish to be able to perform a NNLO analysis, in which these
uncertainties become smaller. For this, the 3--loop splitting
functions have yet to be calculated.
\section*{Direct Photons}
\noindent
A comparison of direct photon data from a number of experiments to
NLO QCD predictions reveals
\cite{S1:huston,S1:marek,S1:cteqdp}
a pattern of systematic deviations that cannot be reproduced by any
modification to the parton distributions, see  Figure~7.
One possible explanation for the
deviations is that they are due to soft gluon emissions, i.e. ${k}_{T }$
effects, similar to those encountered in Drell-Yan production.
In the latter case, a complete description of the
soft gluon effects requires a resummation-type
calculation.

The effects can be approximated using a Gaussian smearing procedure
and/or
parton shower Monte Carlo programs. Soft gluon emission steepens the
direct photon cross sections at low transverse momentum
and, in the case of fixed target
direct photon experiments, changes the normalization. The value of the
${k}_{T}$
can be measured directly in diphoton production, as in the Drell-Yan
process,
and by measuring the properties of the away-side jet in direct photon
production~\cite{S1:marek}.
The theoretical expectation is that the average value
of the ${k}_{T}$ should increase roughly
logarithmically with the center of mass
energy.

A complete description of direct photon data, especially at fixed
target energies, may await a complete resummation calculation of the
cross
sections. Preliminary work in that direction has already taken place
\cite{S1:cp}. As an intermediate step, Baer and Reno \cite{S1:baer}
have added parton showering
to a NLO photon production calculation and have shown that it is possible to
describe the rise observed at low ${p}_{T}$ in CDF at both
$\sqrt{s}$ = 1800 GeV and $\sqrt{s}$ = 630 GeV~\cite{S1:blair}.
%
%
%

Previous common wisdom held that fixed target direct photon data
served to fix the gluon distribution in the proton, since the dominant
mechanism for prompt photon production  in proton-proton collisions
is gluon-quark
scattering. However, the ${k}_{T}$ effects mentioned above, and the
residual
scale-dependence at NLO of the direct photon theory, can affect
both the slope and normalizations of the cross sections.

An attempt has been made to incorporate the fixed target direct
photon data from the
Fermilab  experiment E706~\cite{S1:marek}
into
the CTEQ global fitting program, first correcting for the soft-gluon
(${k}_{T}$)
effects. The direct photon data probe the gluon distribution up to
very high values of $x \sim 0.7$. The gluon distribution obtained from the
fit
agrees well with that from CTEQ4M.

\noindent
\begin{center}
\mbox{\epsfig{file=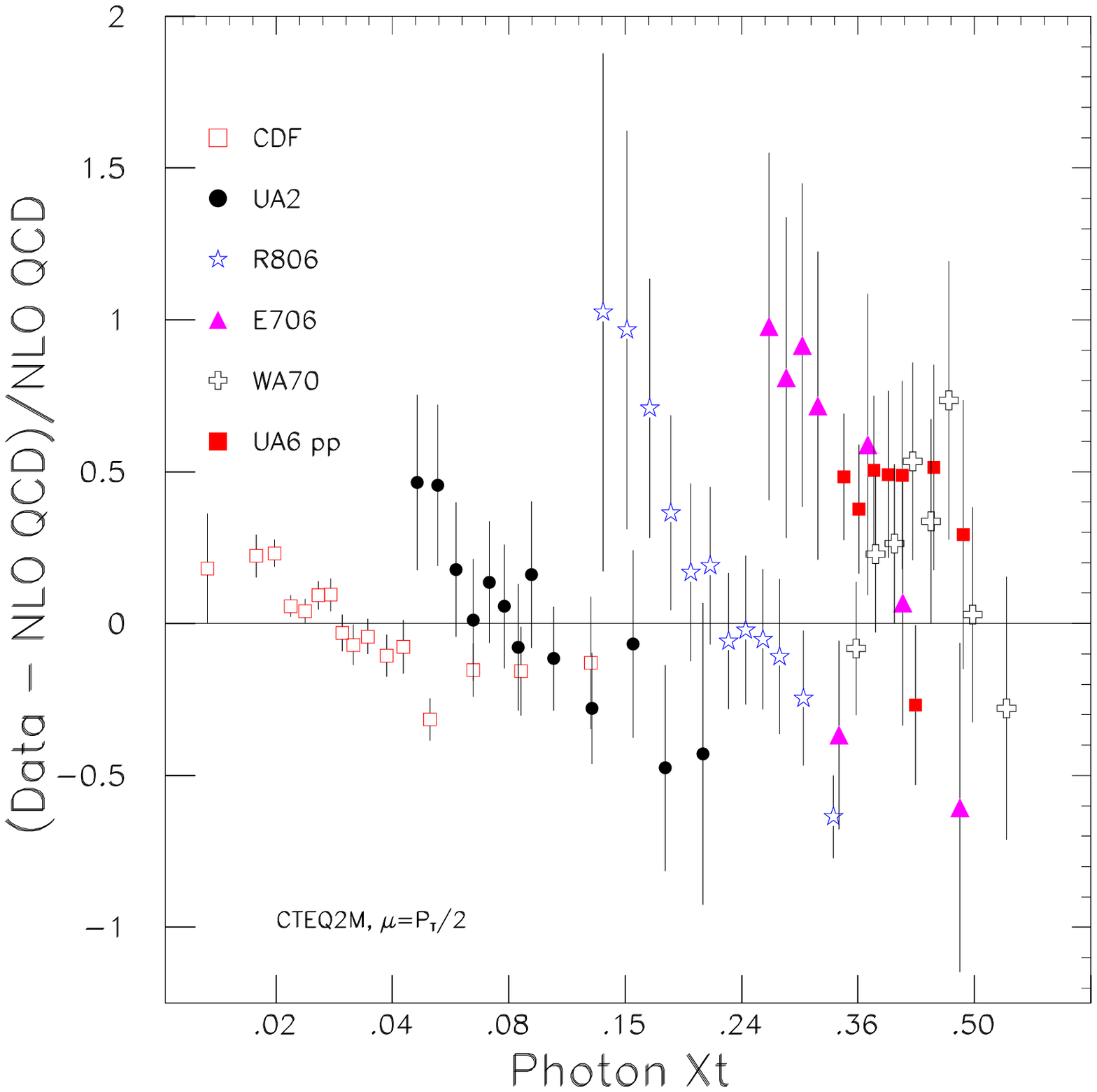,height=8.cm,width=11cm}}
\vspace{2mm}
\noindent
\end{center}

\vspace{-7mm}
\noindent
{\sf Figure~7:~The quantity (Data-Theory)/Theory plotted vs ${x}_{T}$
for a number of direct photon experiments. Theory corresponds to
NLO order QCD with CTEQ2M parton distribution functions.}

\vspace{2mm}

\noindent
 The average ${k}_{T}$ values used
(1.15 GeV/$c$ at $\sqrt{s}$ = 31.5 and 1.30 GeV/$c$ at 38.7 GeV)
agree with the ${k}_{T}$ values
measured in Drell-Yan experiments in similar kinematic regimes and with
the direct determination of the event ${k}_{T}$ from the E706 data.
The ${k}_{T}$-corrected
cross sections describe well the slope and the normalization of the data,
even at the highest values of $x$.

\section*{The High    \mbox{\boldmath{$Q^2$}} Range}

\noindent
\begin{center}
{\large\sf Jet Rates at the Tevatron}
\end{center}

\vspace*{3mm}
\noindent
The jet production at the Tevatron  serves as a crucial venue
for the measurement of the gluon distribution,
the determination of ${\alpha}_{s}$, and probing of parton distributions
at highest $x$ and ${Q}^{2}$ values \cite{S1:huston}.

The leading order cross sections for jet production are proportional
to ${\alpha}_{s}^{2}$, and to ${G}^{2}(x,Q)$ for gluon-gluon
scattering, $G(x,Q)Q(x,Q)$ for gluon-quark scattering, and
$Q(x,Q)Q(x,Q)$  for quark-quark scattering. In the mid-range of the
inclusive
central jet data, $50 \GeV   < E_{T} < 200 \GeV  $, $gg$ and $gq$
scattering dominate.
This is a region where systematic errors, both theory and experiment,
are
smallest and where the theory agrees well with the data. The CDF and D0
cross
sections are also within relatively good agreement within this range
\cite{S1:huston,S1:frank}. At
high values of ${E}_{T}$, CDF has reported an excess
over standard theoretical predictions.

The CDF \cite{S1:cdfjet} and D0 \cite{S1:d0jet} jet data have been
included in a global parton
distribution analysis (CTEQ4M) \cite{S1:cteq4}.
The jet inclusive data serve to stabilize
the fits and to constrain the gluon distribution in the $x$
range from  0.05  to 0.22.
The inclusive jet data prefer a value of ${\alpha}_{s}$ of
0.116 to 0.118, consistent with the global fit results.

As mentioned previously, CDF has reported an excess in the inclusive
cross section at the highest values of transverse energy. This may be
due to
new physics or to a possible modification of the parton
distributions~\cite{S1:cteq4hj}. The
quark distributions in the $x$
and ${Q}^{2}$ range $(0.4-0.5, 1-2 \times {10}^{5} \GeV^2)$
corresponding to high ${E}_{T}$ central jet production are
fairly well-constrained,
but there is greater flexibility and uncertainty in the gluon
distribution.
This freedom is largely due to two sources: the theoretical
uncertainties
in fixed target direct photon production and the lack of other
data in current global
fits to constrain the gluon distribution at high $x$.

\begin{center}
\mbox{\epsfig{file=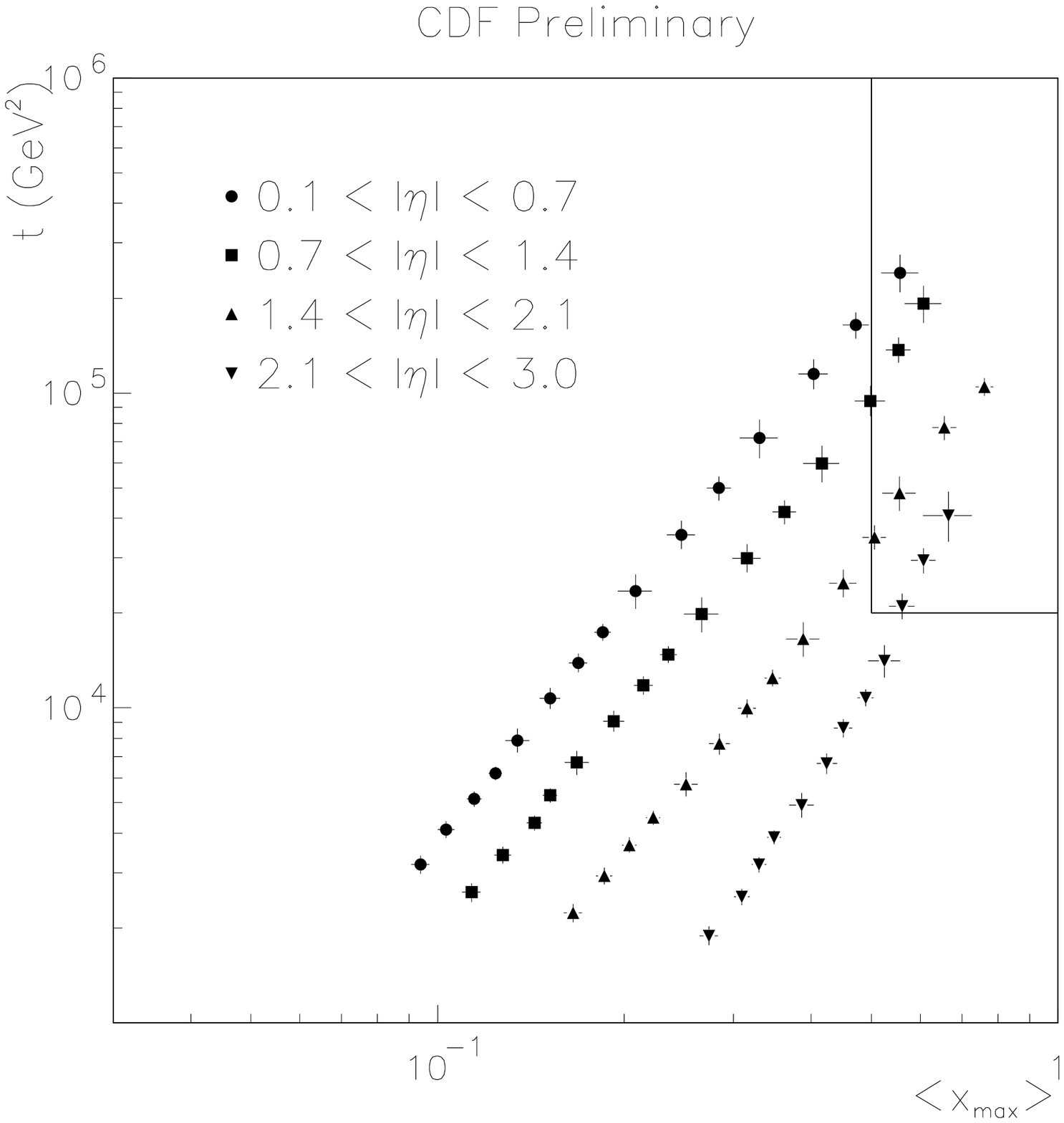,height=8.9cm,width=11cm}}
\vspace{2mm}
\noindent
\end{center}

\vspace{-7mm}
\noindent
{\sf Figure~8:~The momentum-transfer-squared $(t)$ plotted as a function
of the larger of the $x$  values in the parton-parton collision for the
CDF differential dijet cross section. The box indicates the region of the
HERA excess.}

\vspace{2mm}
\noindent
One cannot
conclude that
the excess is due to gluons, just that the data currently used
 in the
global fits
allow for a description of the excess by a larger gluon distribution.
No
definite conclusions can be drawn about the larger gluon distribution
without the use of additional data.
Note also that since
the dominant jet production mechanism in this ${E}_{T}$ range is
quark-quark
scattering, any appreciable modification to the jet cross section
requires
a much larger modification of the gluon distribution.

This exercise with the high ${E}_{T}$ jet distribution also points out
the need for flexibility in the
parameterization of parton distributions, especially when extrapolating
to new regions of $x$ and ${Q}^{2}$ \cite{S1:huston,S1:SF?}.

One can specifically probe high values of $x$  by measuring the
differential dijet cross section. In the CDF differential dijet analysis
\cite{S1:frank}, one jet (the trigger jet) is required to be central
(0.1 ${<}|{\eta}|{<}$ 0.7)
while the other jet (the probe jet) is allowed to have
an $|{\eta|}$ value of up to 3. The trigger and probe jets are the two
highest
${E}_{T}$ jets in the event.

High ${E}_{T}$ scatters in which the probe jet is
at high rapidity correspond
to collisions of a parton at moderate $x$  (0.05--0.3) with a high $x$
parton (up to $x \sim 0.75$)
at  very high ${Q}^{2}$ (up to  ${10}^{5} \GeV^{2}$).
The $x$  and ${Q}^{2}$ values probed
overlap (and in some cases exceed) those in the HERA high $x$ and ${Q}^{2}$
analysis, as shown in Figure~8.
Any modifications to the parton distributions to explain the HERA
high $x$  and ${Q}^{2}$ results \cite{S1:SF?} would have a similar impact
on the differential dijet data. The data should also serve to test the
hypothesis of a larger gluon at high $x$.
One complication is that for large trigger jet ${E}_{T}$ values
and  high probe jet rapidities, the cross section is dominated by
multijet final states due to the larger phase space.
One either needs to compare to higher order calculations or to
measure the cross sections as a function of the parton light-cone
momentum fractions ${X}_{A}$ and ${X}_{B}$, and the dijet pseudorapidity
difference ${\eta}$*, as suggested by Ellis and Soper~\cite{S1:Soper}.
A measurement of this type should be less sensitive to
higher order corrections and should provide a more definitive statement
on the parton distributions at high $x$ and ${Q}^{2}$.

\vspace*{3mm}
\noindent
\begin{center}
{\large\sf Event Rates at HERA}
\end{center}

\vspace*{3mm}
\noindent
The ZEUS and H1 experiments have reported on an excess of events
above standard model predictions at high $Q^2$ and large values of
$x$ \cite{S1:carli,S1:zarnecki}. With a
total luminosity of $14~pb^{-1}$, H1 found 12 events
above $Q^2$ of 15000 GeV$^2$ where 4.7 events are expected.  ZEUS
found 4 events for $x >$ 0.55 and $y >$ 0.25
where 0.9 events are expected.
A quantitative comparison of the distributions from
the two experiments is not possible since the
resolution effects are not
the same for the two data sets.
The uncertainties in the Standard Model predictions are dominated
by the uncertainties in the structure functions, measured at fixed
target experiments and evolved to high $Q^2$.  This uncertainty is
at the level of 5--7\%~\cite{S1:botje,S1:carli,S1:zarnecki} in case that
the conventional parametrizations of the parton distributions are used.

The H1 collaboration has determined the probability that there is
a statistical fluctuation of the same or larger magnitude as observed
at $Q^2  >$ 15000 GeV$^2$ to amount 0.006.  Similarly the
ZEUS collaboration finds  this probability for
the region $x >$ 0.55 and $y >$ 0.25 to be 0.072.
\pagebreak
\vspace*{3mm}
\noindent
\begin{center}
{\large\sf Phenomenological Aspects}
\end{center}

\vspace*{3mm}
\noindent
Shortly after the publication of the excess of events in the high $Q^2$
range by both the H1 and ZEUS experiments~\cite{S1:carli,S1:zarnecki}
a series of
theoretical and phenomenological studies was carried out seeking
for possible interpretations~\cite{S1:REF}. The observed excess was
neither predicted nor do we have  a thorough theoretical concept
to explain it currently.
During this workshop interpretations as due to single
leptoquark production~\cite{S1:JBL1},
effects due to $R$-parity violating supersymmetry
\cite{S1:RPAR}, the presence of contact terms
\cite{S1:CONT}, and others \cite{S1:ARW}, as well as
uncertainties in the knowledge of the structure
functions at larger values of $x$ \cite{S1:SF?} and
consequences for $e^+e^-$ experiments \cite{S1:EPEM} were
discussed. The Tevatron experiments reported on constraints which
have been derived by them for the leptoquark pair production cross
section \cite{S1:DAT}.

\vspace*{3mm}
\noindent
\begin{center}
{\large\sf Leptoquarks}
\end{center}

\vspace*{2mm}
\noindent
If the observed high-$Q^2$ excess is interpreted in terms of single
leptoquark production constraints on the fermionic couplings $\lambda$
of the leptoquarks $\Phi$, which may be either scalars or vectors,
may be derived. Due to the location of the excess found by H1 in the
range $M = \sqrt{x S}  \sim 200~\GeV$ we  assume this scale
in some of the estimates being considered below.
In the narrow width approximation the production cross
section reads
\begin{equation}
\sigma = \frac{\pi^2}{2} \alpha \left ( \frac{\lambda}{e} \right )^2
q(x, \langle Q^2 \rangle) \left \{\begin{array}{c} 2~:~V\\ 1~:~S
\end{array} \right. \times Br(\Phi \rightarrow e q)~.
\end{equation}
For the observed excess in the $e^+ + jet$ channel at H1
\begin{equation}
\label{S1:E:coup}
\frac{\lambda_S}{e} \sqrt{Br}
\sim 0.075~~(0.15)~~~~u~~(d),~~~\Phi = S
\end{equation}
is derived, cf.~\cite{S1:JBL1,S1:JBL2}\footnote{Earlier references on
the phenomenology of leptoquarks may be found in
\cite{S1:JBL1,S1:JBL2,S1:BB}.},
while $\lambda_V = \lambda_S/\sqrt{2}$
and $\lambda_{\rm ZEUS} = 0.55 \lambda_{\rm H1}$. These couplings are
well compatible with the limits derived from low energy
data~\cite{S1:LE}.  An information on the spin of the produced state
can be derived from the $y$ distribution of the events. The statistics
is yet to low to allow for such a detailed
analysis, however, one may compare the average values. For the H1
events one obtains
$\langle
y \rangle_{\rm H1} = 0.59 \pm 0.02$, which is compatible with both the
expectation for a scalar $\langle y \rangle_S = 0.65$ or a vector
$\langle y \rangle_V = 0.55$, cf.~\cite{S1:JBL2}.

A severe constraint on the leptoquark states which may be produced
in $e^+ q$ scattering is imposed by their
    $SU(2)_L \times U(1)_Y$ quantum
numbers\footnote{For a classification in the case of family-diagonal,
baryon- and lepton number conserving, non-derivative couplings, see
\cite{S1:BRW}.}.  If besides the $e^+ q$ final states the indication
of also $\nu q$ final states becomes manifest, no scalar leptoquarks
are allowed since low energy constraints demand either
$\lambda_L \ll \lambda_R$ or $\lambda_R \ll \lambda_L$.  In this case
only the vectors  $U^0_{3\mu}$ or $U_{1\mu}$, which may be produced
in the $e^+ d$ channel,  are allowed.
 For these states the branching ratios
are $Br(e^+ d) = Br(\nu u) = 1/2$. If the observed excess turns out
to be due to single leptoquark production also $e(\nu) + 2 jet$
final states have to be observed with $M_{e(\nu) + jet} \sim 200 \GeV$
\cite{S1:BK}.

Experimental constraints on the existence of leptoquarks in the mass
range $M \sim 200 \GeV$ can be derived from the Tevatron data for the
leptoquark pair production cross section~\cite{S1:BBK}. Results of this
analysis were reported in~\cite{S1:DAT}. The current bounds\footnote{
At the time of the workshop the bounds were lower by 35 GeV.} for
leptoquarks associated to the 1st fermion family
are~\cite{S1:DAT}~:
\begin{eqnarray}
\label{S1:E:bou}
M_S &<& 210~\GeV,~~~~~~~~~~~~~Br(eq) = 1~~~~~~~~~~~(95\%~CL)\nonumber\\
M_S &<& 190~(225~\GeV),~~~~~Br(eq) = 0.5~(1)~~~~(95\%~CL).
\end{eqnarray}
The corresponding limits for vector leptoquarks are expected to be
larger~\cite{S1:BBK}, since the pair production cross section is
larger by a factor of at least $2...3$ for a factorization mass of
$\mu = M_{\Phi}$~\cite{S1:JBL2}.
Whereas leptoquarks with $Br(eq) = 1$ are ruled out at $95 \%~CL$ by
the Tevatron bounds, those with $Br(eq) = 0.5$ are still (marginally)
allowed.

\vspace*{3mm}
\noindent
\begin{center}
{\large\sf R-parity violating SUSY}
\end{center}

\vspace*{2mm}
\noindent
Supersymmetric theories with $R$-parity breaking  bear also scalar
leptoquark states~\cite{S1:RPAR}. The tight bounds from Tevatron
for states with $Br(eq) =1$ do not apply to this class of leptoquarks
since the cascade decays contain a light (unvisible) supersymmetric
particle in the final state. Among the different possible states
$\tilde{u}_L$ is excluded due to limits from $\beta\beta$ decay.
$\tilde{c}_L$ is allowed and would lead to a contribution
to $Br(K^+ \rightarrow \pi^+ \nu \overline{\nu})$ close to the current
bound on this process. Both the processes
$e^+ d \rightarrow \tilde{t}_L$ and $e^+ s \rightarrow \tilde{t}_L$
are allowed at a sufficient production rate.

\vspace*{3mm}
\noindent
\begin{center}
{\large\sf Contact Interactions}
\end{center}

\vspace*{2mm}
\noindent
Since part of the current high $Q^2$ events are distributed in a
somewhat wider range one might as well try to describe them with the help
of an effective Lagrangian containing contact interactions at some
scale $\Lambda$,
\begin{equation}
\label{S1:E:conta}
L_{eff} = \sum_{q = u,d} \sum_{h_1, h_2 = L,R} \eta^{eq}_{h_1, h_2}
\overline{e}_{h_1} \gamma^{\mu} e_{h_1} \overline{q}_{h_2} \gamma_{\mu}
q_{h_2},
\end{equation}
with
\begin{equation}
\eta^{eq}_{h_1, h_2}  = \pm \frac{4 \pi}{\Lambda^2_{h_1, h_2}},~~~~~~~
[\Lambda] = \GeV~.
\end{equation}
Various of the possible interactions described by (\ref{S1:E:conta})
are already well constrained.
$e^+p$ interactions are particularly sensitive to
the $\eta_{LR}$ and $\eta_{RL}$ terms~\cite{S1:CONT}, which could provide
a description with $\Lambda$ being of $O(3~\TeV)$.

\vspace*{3mm}
\noindent
\begin{center}
{\large\sf Uncertainties of parton densities at larger $x$}
\end{center}

\vspace*{2mm}
\noindent
Both the H1 and ZEUS collaborations estimated the uncertainty of the
quark densities in the      range $x \sim 0.4 ... 0.5$, in which the
excess of events is observed  to be of $O(5-7\%)$. As well--known, the
parton densities are not yet well constrained
experimentally at very
large $x$, $x \gsim 0.8$. A feed--back of this uncertainty into the
range of lower $x$ values in QCD fits is not excluded and was studied
in \cite{S1:SF?} under a series of assumptions. Although more studies
are needed the above mentioned uncertainty is essentially confirmed
and the excess of events cannot be attributed to an uncertainty in the
parton densities currently.

\vspace*{3mm}
\noindent
\begin{center}
{\large\sf Implications for $e^+e^-$ scattering}
\end{center}

\vspace*{2mm}
\noindent
The fermionic couplings of leptoquark states in the mass region
$M \sim 200 \GeV$ may be probed in $e^+e^-$ scattering searching for a
possible interference of leptoquark exchange in the $t$-channel
and $\gamma-Z$ exchange in the $s$-channel.
The corresponding effects at SLC, LEP1 and LEP2, however, are
smaller than 1\%~\cite{S1:NG,S1:EPEM} assuming the
couplings~(\ref{S1:E:coup}) and can thus
not be detected for leptoquarks in this  mass range.
On the other hand, LEP2 may constrain related contact interaction
terms up to scales of $\Lambda \gsim 4 ... 6.5 \TeV$~\cite{S1:EPEM}
 on the basis
of the data being accumulated until
 the end of this year. Furthermore
one may as well search for some specific signatures associated to
leptoquarks which emerge in scenarios with supersymmetric models
with $R$--parity violation.
\section*{Conclusions}
\noindent
There has been considerable progress in the measurement of the deep
inelastic scattering structure functions during the last year. With
the increased luminosity, which will be collected by the HERA experiments
until DIS~'98, the precision to  which
$F_2(x,Q^2)$ and $F_2^{c\overline{c}}$ will be determined will be
higher and  allow for even more detailed comparisons with the
theoretical predictions. With the availability of the complete
small $x$ resummed anomalous dimension $\gamma_{gg}$ a more complete
analysis of the small $x$ behavior  of $F_2(x,Q^2)$ can be performed.
To be able to analyze also the behavior  of the structure functions for
lower values of $Q^2$ the twist--4 anomalous dimensions
need to be known.

The most exciting question is certainly whether or not the excess of
neutral current events in the high $Q^2$ region of HERA will persist
adding in the data of the 1997 runs, and whether such an excess is also
present in the charged current data. This question will be answered
before DIS~'98 by the HERA experiments. If a clear manifestation of new
physics will be found, will we have the theory for it in
a year from now?

\vspace{2mm}
\noindent
{\bf Acknowledgement.} We would like to thank all the speakers for their
contributions to the
working group.  Our thanks are also due to  M. Derrick,
D. Krakauer, J. Repond, and the other members of the organizing
committee of DIS '97 for organizing a very fruitful meeting in a
splendid atmosphere.

\end{document}